\documentclass[preprint,onecolumn,superscriptaddress]{revtex4}

\usepackage{amssymb}
\usepackage{enumerate}
\usepackage{amssymb}
\usepackage{amsmath}
\usepackage{feynmf}
\usepackage{slashed}
\usepackage{natbib}
\usepackage{graphicx}
\usepackage[pdftex,bookmarks,linktocpage,pdfpagelabels,plainpages=false,hyperfigures,linkcolor=blue,citecolor=blue]{hyperref} %for hypertext links
\hypersetup{
    colorlinks = true,
    allcolors = blue,}

\usepackage{appendix}

\usepackage{mathrsfs,amssymb}  
\usepackage{cancel}
\usepackage[normalem]{ulem}

\newcommand\be{\begin{equation}}
\newcommand\ee{\end{equation}}

\newcommand\bea{\begin{eqnarray}}
\newcommand\eea{\end{eqnarray}}

\begin{document}
\bibliographystyle{apsrev4-1}

%\title{Testing Beyond Standard Model Neutrino Interactions at the Coherent CAPTAIN Mills Experiment} 
\title{Sailing the CE$\nu$NS Seas of Non-Standard Neutrino Interactions with the Coherent CAPTAIN Mills Experiment}

\author{Ian M. Shoemaker}
\email{shoemaker@vt.edu}
\affiliation{Center for Neutrino Physics, Department of Physics, Virginia Tech University, Blacksburg, VA 24601, USA}

\author{Eli Welch}
\affiliation{Center for Neutrino Physics, Department of Physics, Virginia Tech University, Blacksburg, VA 24601, USA}

\begin{abstract}

We study future coherent elastic neutrino-nucleus scattering (CE$\nu$NS) modifications from a variety of possible models at the Coherent CAPTAIN Mills (CCM) experiment at Los Alamos. We show that large regions of Non-Standard Neutrino Interaction (NSI) parameter space will be excluded rapidly, and that stringent new bounds on the gauge coupling in $Z'$ models will also be placed. As a result, CCM will be able to rule out LMA-D solutions for a large class of models with MeV-scale mediators. 

%\com{Add discussion about perturbative couplings despite large cross sections...} \\

\end{abstract}

\maketitle

\section{Introduction}

In the aftermath of the observation of neutral currents by Gargamelle in 1973~\cite{Hasert:1973ff}, it was quickly realized that the neutral currents will induce a large coherent cross section on nuclei at low energies~\cite{Freedman:1973yd}. In 2017, this Coherent Elastic Neutrino-Nucleus Scattering (CE$\nu$NS) was observed by the COHERENT collaboration, yielding qualitatively new information on neutrinos~\cite{Akimov:2017ade}.  This is the largest cross section at low-energies, and its utility allows for the study of precision neutrino-nucleus interactions~\cite{Scholberg:2005qs,Coloma:2017egw,Shoemaker:2017lzs,Liao:2017uzy,Denton:2018xmq}, sterile neutrinos~\cite{Anderson:2012pn,Dutta:2015nlo,Kosmas:2017zbh}, as well as non-neutrino BSM such as light DM~\cite{deNiverville:2015mwa,Ge:2017mcq,Dutta:2019nbn,Hurtado:2020vlj,Dutta:2020vop} and ALPs~\cite{Dent:2019ueq,AristizabalSierra:2020rom}.

Looking to the future, the European Spallation Source (ESS)~\cite{Baxter:2019mcx}, the Coherent CAPTAIN-Mills (CCM) at Lujan center at Los Alamos National Laboratory~\cite{CCM},
and Oak Ridge National Laboratory~\cite{Akimov:2019xdj} will have ton and multi-ton detectors deployed. This rapid expansion of statistics with complementarity detector types will allow global CE$\nu$NS data to place powerful new constraints on new physics. Recent work has highlighted the need for a careful statistical treatment~\cite{Denton:2020hop}, the usage of copula in order to combine global data in an effort to break NSI degeneracies~\cite{Dutta:2020che}, and combining timing and energy information to maximize the search sensitivity to new physics~\cite{Coloma:2019mbs,Dutta:2019nbn}. There may be a hint of nonzero NSI in the tension between T2K and NOvA~\cite{Denton:2020uda}. Experimentally, much attention has been paid to the precise value of the quenching at the lowest energies for CsI[Na]~\cite{Dutta:2019nbn,Collar:2019ihs}. This can play an important role in determining the precise sensitivity to BSM, especially for example to low-mass mediators of neutrino-nucleus scattering.

In this work, we focus on the BSM capabilities of the CCM experiment at Los Alamos. In particular, we will study its ability to probe NSI and light mediator $Z'$ models. These scenarios are naturally related to each other given that a sufficiently heavy $Z'$ can be integrated out leaving behind the neutral current four-fermion effective operators of NSI. 

For many decades, theorists have considered the existence of new neutral current interactions between neutrinos and matter~\cite{Wolfenstein:1977ue}. Dubbed Non-Standard neutrino Interactions (NSIs), this new physics can alter neutrino production, detection, and their propagation (see the reviews~Ref.~\cite{Davidson:2003ha,Biggio:2009nt,Farzan:2017xzy,Dev:2019anc}). In this work, we will focus on neutral current NSI described by the dimension-6 effective Lagrangian, 
\be \mathscr{L}_{NSI} \supset 2 \sqrt{2} G_{F} \epsilon_{\alpha \beta} \left(\bar{\nu}^{\alpha} \gamma^{\mu} \nu_{\beta} \right) \left(\bar{f} \gamma^{\mu} f \right),
\label{LNSI}
\ee
 where $\alpha,\beta$ are neutrino flavor indices and $f$, is a charged SM fermion, $f = \{e, u,d\}$. The strength of these new operators is parameterized by the $\epsilon$ coefficients in units of $G_{F}$. 
 In the language of EFT, we regard the above four-fermion description as an effective description below the energy scale $\Lambda \equiv 1/\sqrt{2 \sqrt{2} \epsilon G_{F}}$. Historically NSI studies have focused on TeV (or above) scales for the source of new physics giving rise to Eq. (\ref{LNSI}).  In this case, large $\epsilon$'s come about from order unity couplings with EW scale masses.  Given that the impact of such non-renormalizable operators grows rapidly with energy, colliders have placed strong bounds on NSI~\cite{Berezhiani:2001rs,Davidson:2011kr,Friedland:2011za,Franzosi:2015wha,Babu:2019mfe}. 

Some of the most model-independent bounds on NSI come from oscillation data. In this case, NSI modifies the forward coherent scattering of neutrinos as they travel through matter~\cite{Wolfenstein:1977ue}. As such, the presence of NSI impacts the flavor evolution of neutrinos in matter, resulting in tight constraints on NSI~\cite{Farzan:2017xzy,Esteban:2018ppq}, and their presence can significantly complicate the interpretation of oscillation data (see e.g.~\cite{Friedland:2004pp,Friedland:2012tq,Capozzi:2019iqn}).  Bounds of this type apply to most models so long as the mediator is sufficiently heavy ($\gtrsim 10^{-12}~{\rm eV}$)~\cite{GonzalezGarcia:2006vp,Coloma:2020gfv}.

%As a result of the lack of the long-predicted upturn in electron neutrino survival probability, the robustness of the LMA solution...  \com{LMAD}\cite{Miranda:2004nb}

CE$\nu$NS data occupies an intermediate regime of momentum transfer where the model dependence of realistic UV completions of NSI can be probed~\cite{Shoemaker:2017lzs,Liao:2017uzy,Denton:2018xmq}. Although CE$\nu$NS data therefore probes NSI in a less model-independent way than oscillations, this dependence on the underlying physics allows for potential sensitivity to directly measuring the new physics scale associated with NSI~\cite{Shoemaker:2017lzs}. 

The remainder of this paper is organized as follows. In Sec.~\ref{sec:CCM} we review the capabilities of CCM and the assumptions we make regarding its performance. In Sec.~\ref{sec:analysis} we outline our calculational framework, and the assumed BSM modifications to CE$\nu$NS. We find that with CCM alone, diagonal NSI constraints will be greatly improved, although some degeneracies will remain. We also find that simple $Z'$ models will be ruled out as possible LMA-D solutions~\cite{Miranda:2004nb}, highlighting the important complementarity between CE$\nu$NS and oscillation data. Finally in Sec.~\ref{sec:disc} we discuss future directions and conclude in Sec.~\ref{sec:conclusions}.

\section{CCM Capabilities}
\label{sec:CCM}

A significant advance in CE$\nu$NS data will happen soon, thanks to the Coherent CAPTAIN (Cryogenic Apparatus for Precision Tests of Argon Interactions with Neutrinos) Mills (CCM) experiment. The experiment will take place at Los Alamos National Lab's  Lujan Center at the Los Alamos Neutron Science Center (LANSCE). A very high instantaneous power is crucial for measuring the signal-to-background ratio, and makes the Lujan Center a competitive environment for neutrino physics. While a 300 ns beam pulse width can reduce background by a factor $\sim 10^{5}$, they expect to reach 30 ns widths with an order of magnitude gain in background rejection.   

%%%%%%%%
\begin{figure*}[t!]
\includegraphics[angle=0,width=.48\textwidth]{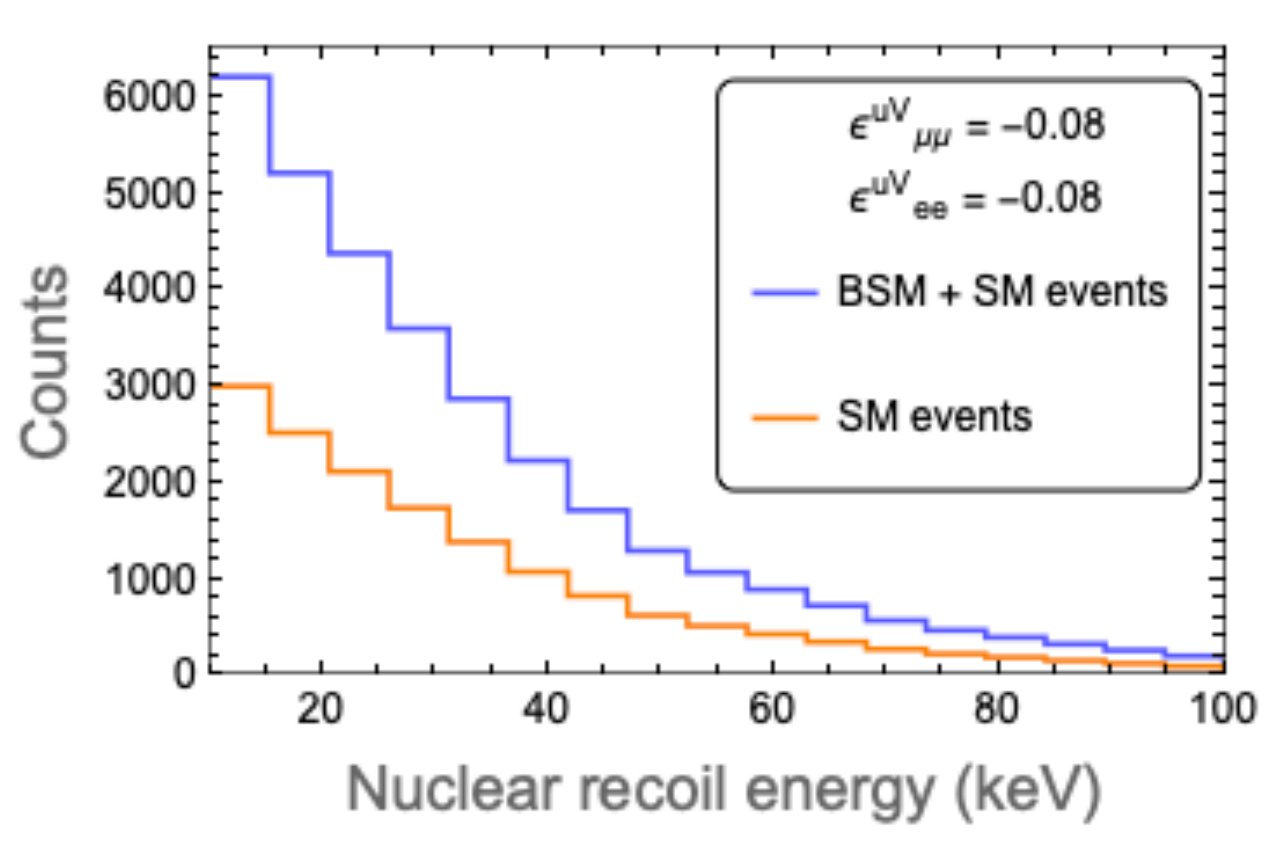}%
\includegraphics[angle=0,width=.48\textwidth]{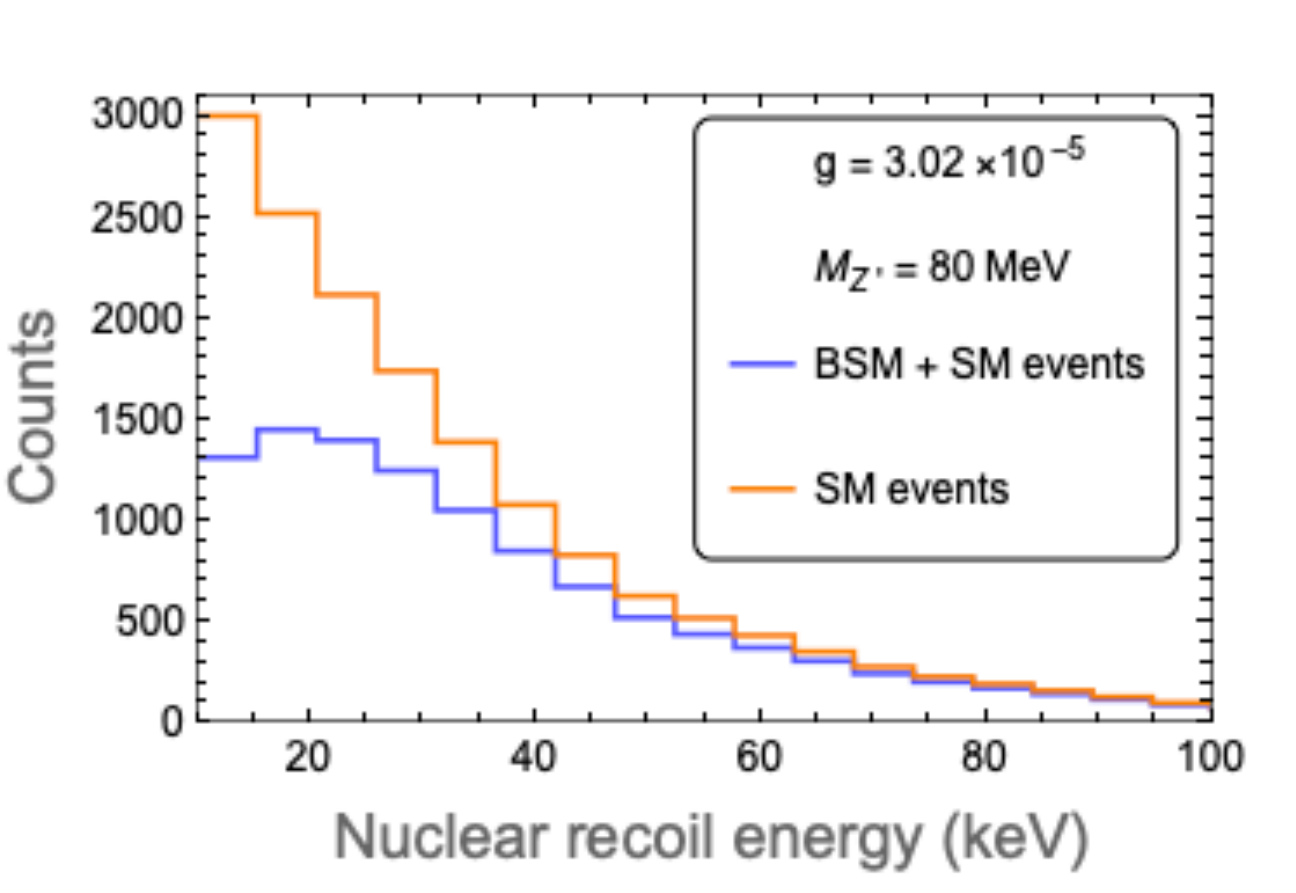}%
   \caption{Illustrative predicted event spectra at CCM for heavy mediator NSI (left) and light mediator $Z'$ (right).}
   \label{fig:spectra}
\end{figure*}
%%%%%%%%

Following~\cite{CCM} we assume that the CCM will employ a fiducial 7 ton liquid argon (LAr) detector with 3 tons of LAr veto. The active veto will be employed to reject cosmic ray and other relevant backgrounds. Assuming a 25 keV threshold, 20 meters from the target, the neutrino flux at the detector will be $\sim 4.5 \times 10^{5}~{\rm cm}^{-2}~{\rm s}^{-1}$~\cite{CCM}. In what follows we will assume 3 years of running. 

\section{CCM Analysis Setup}
\label{sec:analysis}

We will assume that the neutrino flux is well-described by pion decay at-rest, $\pi^{+} \rightarrow \mu^{+} + \bar{\nu}_{\mu}$, followed by $\mu^{+} \rightarrow \nu_{\mu} + e^{-} + \bar{\nu}_{e}$. This induces a sharply peaked muon anti-neutrino flux, resulting from two-body decay kinematics. In contrast, the electron anti-neutrino and muon neutrino arising from muon decay are distributed over a range of energies up to a cut-off around 50 MeV. As such we model the fluxes as:
\be
\phi_{\nu_\mu}(E_\nu)=\phi_0~\delta(E_\nu-E_{\nu_0})
\label{eq:flux1}
\ee
where $ E_{\nu_0}\equiv \frac{m^2_\pi -m^2_\mu}{2m_\pi} \approx 29.8$ MeV. And  
\be
\phi_{\bar\nu_\mu}(E_\nu) = \phi_0~ \frac{64E^2_\nu}{m_\mu^3}~\left(\frac{3}{4}-\frac{E_\nu}{m_\mu} \right),
\label{eq:flux2}
\ee
\be
\phi_{\nu_e}(E_\nu) = \phi_0~ 
\frac{192E^2_\nu}{m_\mu^3}~\left(\frac{1}{2}-\frac{E_\nu}{m_\mu} \right)
\label{eq:flux3}
\ee
where the $\phi_{0}$ normalization is fixed by the number of proton on target and the distance to the detector.

Next we write the CE$\nu$NS differential cross section on a nucleus of mass $M$ as
\be
\frac{d\sigma_{\alpha}}{dE_r} = \frac{G^2_F}{2\pi} Q^2_\alpha F^2(2M E_r)M\left(2-\frac{M E_r}{E^2_\nu} \right)
\label{eq:cross}
\ee
where $E_{r}$ is the nuclear recoil energy, and $F(Q^2)$ is the nuclear form factor, which we parameterize as in~\cite{Engel:1991wq}. The effective charge $Q_{\alpha}$ in the SM is given by
\be
Q^2_{\alpha,SM} = (Zg^V_p+Ng^V_n)^2,
\ee
where $N$ and $Z$ are the number of neutrons and protons respectively, and the SM $Z^0$ couplings to proton and neutron are $g_p^V=\frac{1}{2}-2\sin^2 \theta_w,$ and $    g_n^V=-\frac{1}{2}$.

More generally, we will consider two BSM scenarios which modify the effective charge $Q_{\alpha}$. First we will examine conventional NSI in the heavy mediator limit, as in Eq.~\ref{LNSI}.
Thus in the heavy mediator limit the effective charge entering into the cross section (Eq.~\ref{eq:cross}) becomes.
\be
Q^2_\alpha = [Z(g_p^V+2\epsilon_{\alpha \alpha}^{uV}+\epsilon_{\alpha \alpha}^{dV})+N(g_n^V+\epsilon_{\alpha \alpha}^{uV}+2\epsilon_{\alpha \alpha}^{dV})]^2+\sum_{\beta \neq \alpha}[Z(2\epsilon_{\alpha \beta}^{u V}+\epsilon_{\alpha \beta}^{d V})+N(\epsilon_{\alpha \beta}^{u V}+2\epsilon_{\alpha \beta}^{d V})]^2,
\label{eq:NSI}
\ee
where the first squared term includes those NSI coefficients which interfere with the SM contribution, and the second squared terms include all the off-diagonal $\epsilon_{\alpha \beta}$ which have no SM analog. For simplicity in this paper we will assume flavor-diagonal NSI, leaving flavor off-diagonal terms for future work.

%%%%%%%%
\begin{figure*}[t!]
\includegraphics[angle=0,width=.44\textwidth]{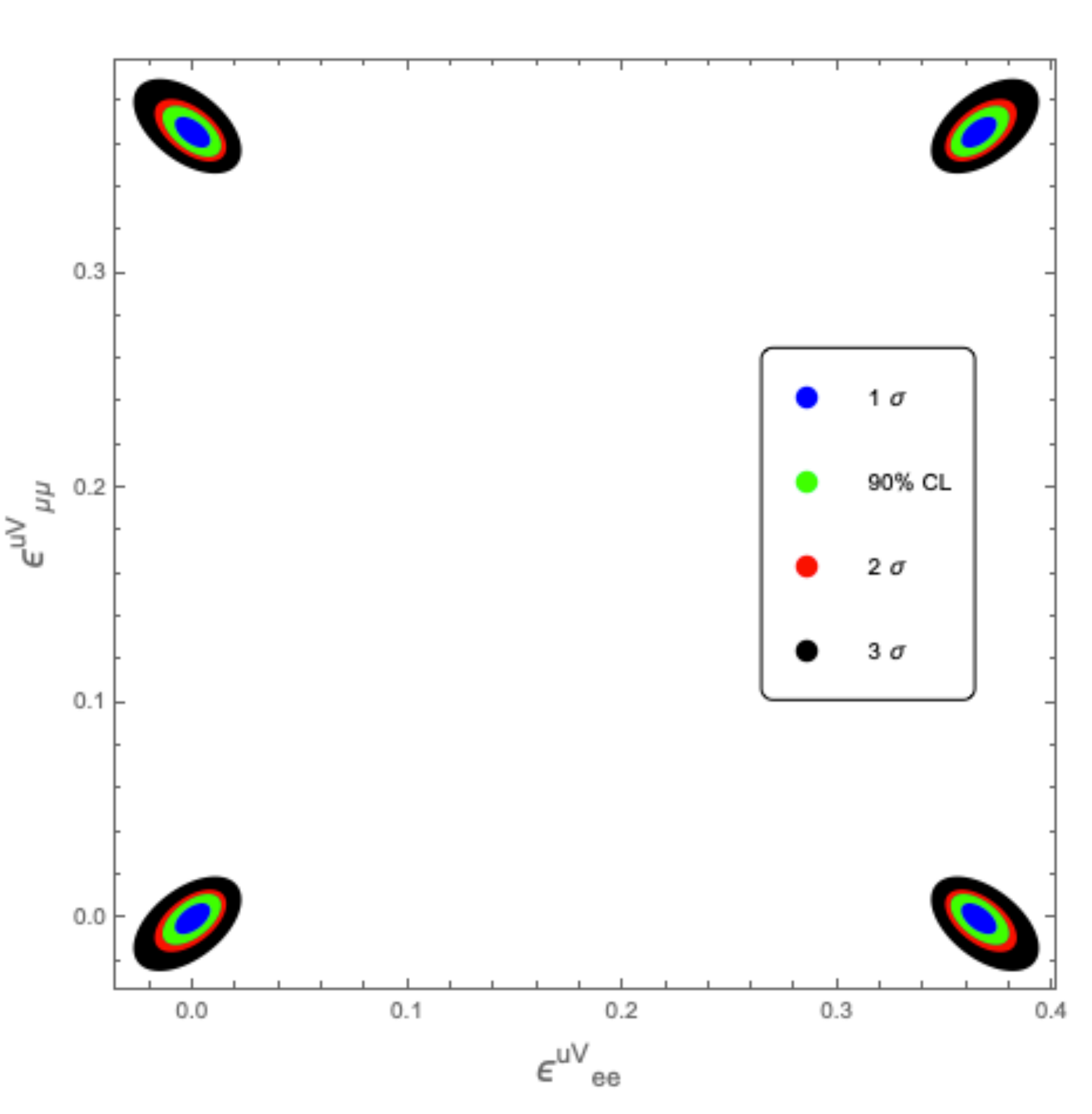}%
\includegraphics[angle=0,width=.51\textwidth]{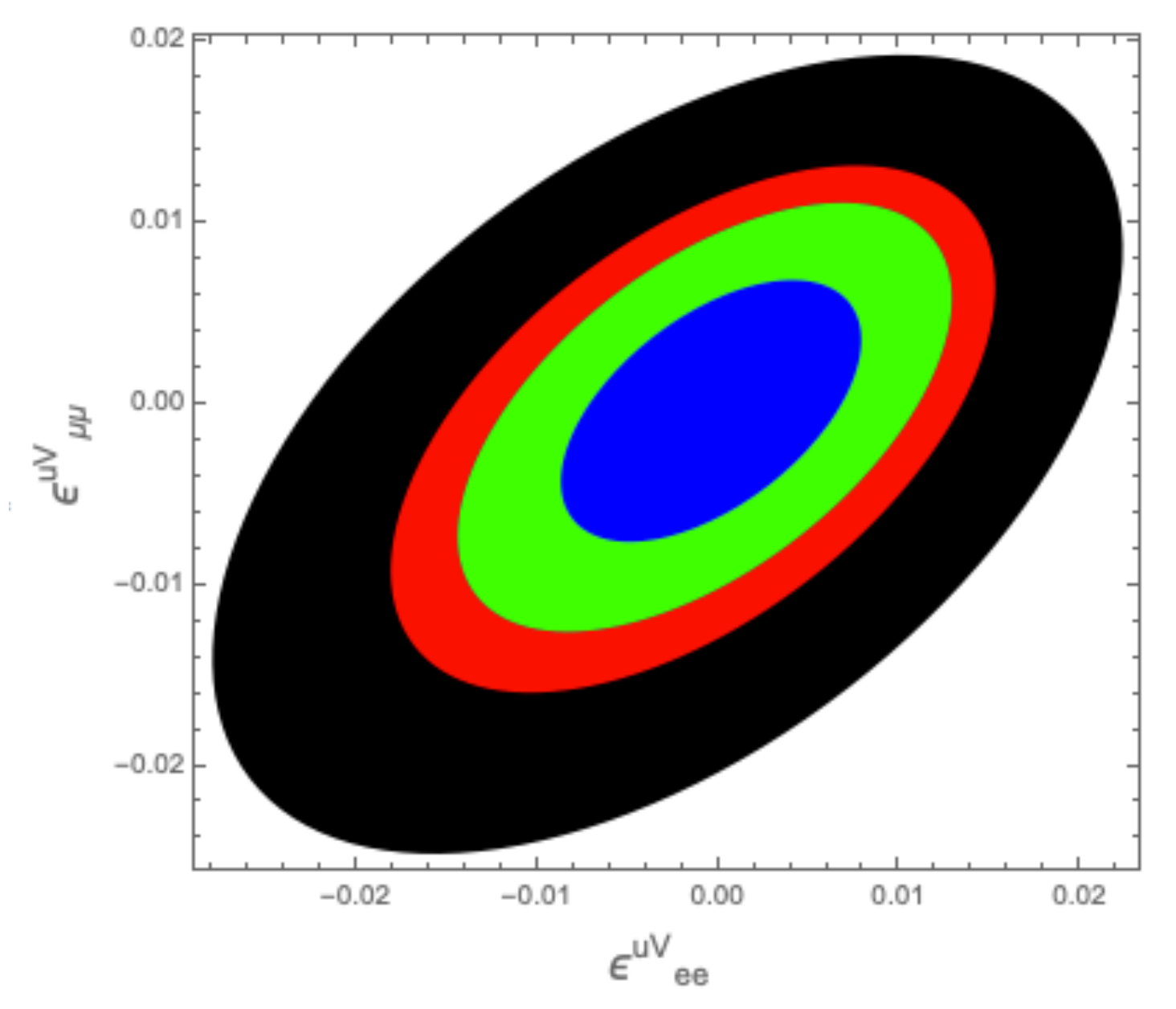}%
   \caption{90$\%$ CL sensitivity projections on NSI parameters for up-quark couplings from the CCM experiment. The constraints on down-type couplings are qualitatively similar. The right panel shows a close-up of the allow region in the immediate vicinity of the SM.}
   \label{fig:NSI}
\end{figure*}
%%%%%%%%

On the other hand, new physics contributions to CE$\nu$NS can be hidden from us at low and accessible mass scales if it is very feebly coupled~\cite{Farzan:2015doa,Farzan:2015hkd,Farzan:2016wym}.  Consider for example the following phenomenological $Z'$ model giving rise to NSI 
\be 
\mathscr{L} \supset g_{q} Z'_{\mu} \bar{q} \gamma^{\mu} q + g_{\nu} Z'_{\mu} \bar{\nu} \gamma^{\mu} \nu
\ee
this yields Eq.(\ref{LNSI}) when the energies of interest are below $ \Lambda \equiv m_{Z'}/\sqrt{g_{\nu} g_{q}}$. For simplicity here we assume that the $Z'$ couples equally to all quark flavors. We will however examine the coupling to electron- and muon-flavor neutrinos separately. With these assumptions the effective charge entering into Eq.~\ref{eq:cross} is
\be
    Q^2_{\alpha}=[Z(g^V_p+\frac{3g^2}{2\sqrt{2}G_F(2ME_r+M^2_{Z'})})+N(g^V_n+\frac{3g^2}{2\sqrt{2}G_F(2ME_r+M^2_{Z'})})]^2
\label{eq:Zprime}
\ee

Finally we compute the differential rate of events per unit detector mass as
\be 
\frac{dR_{\alpha}}{dE_{R}} = \int \phi_{\alpha} \frac{d\sigma_{\alpha}}{dE_{R}} dE_{\nu},
\ee
with $\phi_{\alpha}$ the fluxes given in Eqs.~\ref{eq:flux1},~\ref{eq:flux2}, and \ref{eq:flux3}, and the cross section $d\sigma_{\alpha}/dE_{R}$ given in Eq.~\ref{eq:cross}.

To be conservative we assume that the number of background events will be 20$\%$ of the signal events. Flux normalization uncertainties and detector efficiencies and calibrations will be important systematic uncertainties. As a rough treatment, we take these to be a 10$\%$ overall uncertainty on the flux normalization. These assumptions are conservatively based on COHERENT~\cite{Akimov:2018ghi}, though it is expected that CCM will benefit from an enhanced background rejection capability at the Lujan facility. 

We display the nuclear recoil spectra for both the NSI and $Z'$ models in Fig.~\ref{fig:spectra}. As can be seen, heavy mediator NSI produces a spectrum broadly similar to the SM, though differing of course in normalization. This is expected given that no novel energy dependence arises in heavy mediator NSI (see Eq.~\ref{eq:NSI}). The $Z'$ model on the other hand, includes the explicit energy dependence in Eq.~\ref{eq:Zprime} arising from the $Z'$ propagator. As a result the spectrum appearing in the right panel of Fig.~\ref{fig:spectra} differs in shape compared to the SM expectation. As previously pointed out, such a feature can potentially be used to determine the $Z'$ mass in the event of a discovery~\cite{Shoemaker:2017lzs}.

Next we compute the sensitivity CCM can expect to both heavy mediator NSI and light $Z'$ models. In Fig.~\ref{fig:NSI} we examine the sensitivity to the diagonal elements $\epsilon_{ee}$ and $\epsilon_{\mu \mu}$. Unlike the off-diagonal NSI coefficients, these benefit from interference with the SM CE$\nu$NS contributions and are as a result more tightly constrained. We see that there are four allowed regions, one of which (bottom left) which encompass the SM ($\epsilon=0$). We note that although the regions we find are quite different than those found for COHERENT~\cite{Liao:2017uzy}, they are similar to those found for a projection study of COHERENT before the experiment ran~\cite{Coloma:2017egw}. We note that the apparent NSI sensitivity at CCM will be sufficiently strong to rule out the LMA-D region (see e.g. Fig. 7 from Ref.~\cite{Coloma:2017egw}), which is off the plot in our Fig.~\ref{fig:NSI}. 

%%%%%%%%
\begin{figure*}[t!]
\includegraphics[angle=0,width=.48\textwidth]{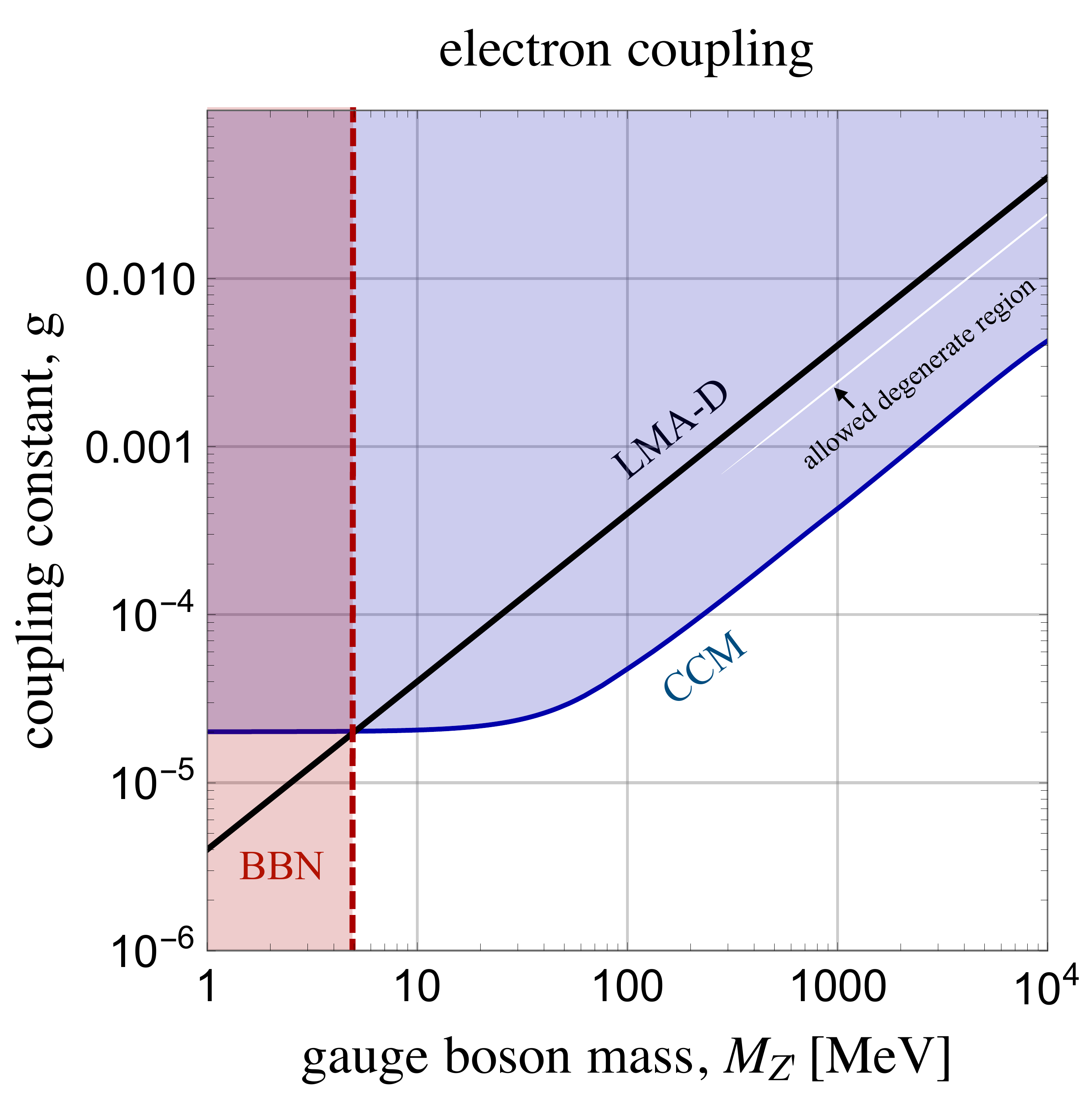}%
\includegraphics[angle=0,width=.48\textwidth]{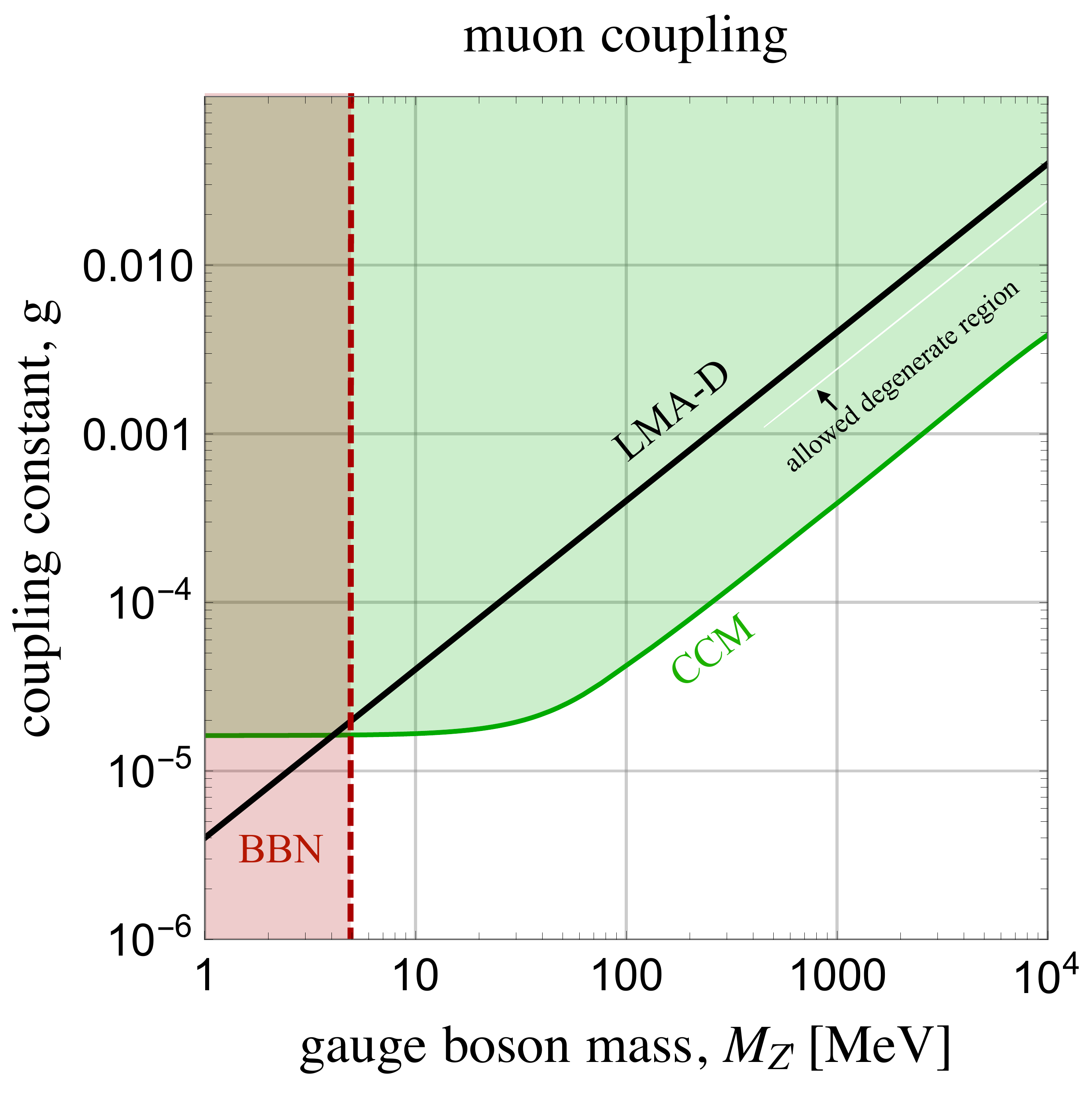}%
   \caption{90$\%$ CL constraints on $Z'$ completions of NSI for electron neutrino couplings (left) and muon neutrino couplings (right). Note the small allowed strip of parameter space in each case (see text for discussion). For reference we also display the BBN~\cite{Kamada:2015era} exclusions and LMA-D region.      }
   \label{fig:zprime}
\end{figure*}
%%%%%%%%

Finally, we display the expected sensitivity to the $Z'$ model in Fig.~\ref{fig:zprime} with only the electron or muon coupling turned on in each panel. We have also included the LMA-D region, and the bounds from BBN which are strong at low mediator masses~\cite{Kamada:2015era}. We have focused on $Z'$ models with equal up and down-quark couplings, but note that it has recently been argued that more general scenarios may remove some of the tension between LMA-D and CE$\nu$NS data~\cite{Esteban:2018ppq,Coloma:2019mbs,EstevesChaves:2021jct}. 

%muon coupling:
%\begin{equation*}
%Q^2_\mu = [Z(g_p^V+2\epsilon_{\mu \mu}^{uV})+N(g_p^V+\epsilon_{\mu \mu}^{uV}]^2
%\end{equation*}

%and
%electron coupling:
%\begin{equation*}
%Q^2_e = [Z(g_p^V+2\epsilon_{e e}^{uV})+N(g_p^V+\epsilon_{e e}^{uV})]^2
%\end{equation*}
%\newline

%SM $Z^0$ couplings to proton and neutron:
%\begin{equation}
%    g_p^V=\frac{1}{2}-2sin^2 \Theta_W
%\end{equation}
%
%\begin{equation}
    %g_n^V=-\frac{1}{2}
%\end{equation}
% \newline
% from https://arxiv.org/pdf/1701.04828.pdf
 %\newline

\section{Discussion and Future Directions}
\label{sec:disc}

In the future we plan to study the utility of multiple detectors with different target nuclei to break degeneracies in NSI couplings which arise in light mediator models. Although NSI coupling degeneracies are studied in many scenarios, they seem not to have been studied in light mediator models which effectively lead to momentum-dependent NSI couplings~\cite{Shoemaker:2017lzs}. As illustrated in Fig.~\ref{fig:zprime}, CE$\nu$NS plays an important in the ability to constrain neutrino oscillations in matter (as in the LMA-D solution), and can shed light on light $Z'$ models that impact BBN~\cite{Kamada:2015era}. For example, CCM found evidence of $Z'$ inside the BBN highlighted region of Fig.~\ref{fig:zprime}, it would indicate that a non-trivial cosmological history played a key role in ameliorating tension between the new light degree of freedom at BBN. 

Additionally, the power of CE$\nu$NS to probe models non-minimal sterile neutrino models has not received much attention. For example, non-minimal Sterile Neutrinos have been shown to be a viable explanation of the XENON1T excess~\cite{Bally:2020yid,Shoemaker:2020kji} and may have a connection to the observed Hubble tension~\cite{Berbig:2020wve}. Similar models can also be probed at CE$\nu$NS experiments via neutrino-to-sterile upscattering. Moreover, in sterile neutrino models with enhanced baryon couplings, the HNLs may be directly produced in the target~\cite{Batell:2016zod}, though this has not been studied in light of present CE$\nu$NS data. We plan to return to these topics in future work.

\section{Conclusions} 
    \label{sec:conclusions}

We have examined the near term sensitivity of the Coherent Captain Mills experiment to a variety of BSM neutrino modifications. First we focused on the sensitivity to neutral current contact operators, i.e. NSI. There we found that the high-statistics which CCM will offer gives it a significant opportunity to break strong NSI parameter degeneracies. We have focused on the flavor diagonal NSI couplings, $\epsilon_{ee}$ and $\epsilon_{\mu \mu}$, but we expect CCM to play an important role in constraining the off-diagonal elements $\epsilon_{e\alpha}$ and $\epsilon_{\mu \alpha}$ as well. Then we examined a simplified $Z'$ model of neutrino-nucleus interactions. We found that CCM will significantly improve over COHERENT's bounds on the gauge. As such this will allow it to close off a small window of parameter space which previously allowed for LMA-D to be compatible with BBN~\cite{Denton:2018xmq}. It is likely that a viable LMA-D solution in the scenario of Ref.~\cite{Farzan:2015doa} with MeV-scale mediators will be ruled out by CCM, though the sub-eV scenario cannot be tested at CE$\nu$NS type experiments. Further, viable LMA-D solutions based on purely leptonic gauge couplings offer an alternative route for sufficiently suppressing CE$\nu$NS constraints~\cite{Coloma:2020gfv}. Lastly, a difference in up- and down-quark couplings can allow for destructive interference on particular nuclei, which can also ameliorate the tension between CE$\nu$NS and LMA-D~\cite{EstevesChaves:2021jct}.

\vspace{1cm}

{\bf \emph{Acknowledgements-  }} We are very grateful for discussions with Bhaskar Dutta and Adrian Thompson. This work is supported by the U.S. Department of Energy under the award number DE-SC0020250.

\bibliography{ref}

\end{document}